\documentclass[doublecol]{epl2} 

\usepackage{subfigure}
\begin{document}

\title{Path Integral approach to nonequilibrium potentials in multiplicative Langevin dynamics}

\shorttitle{Path Integral approach to nonequilibrium potentials} 

\author{Daniel G.\ Barci\inst{1} \and  Zochil Gonz\'alez Arenas\inst{2} \and  Miguel Vera Moreno\inst{1}}
\shortauthor{D. G. Barci \etal}

\institute{                    
  \inst{1} Departamento de F{\'\i}sica Te\'orica,
Universidade do Estado do Rio de Janeiro, Rua S\~ao Francisco Xavier 524,
20550-013,  Rio de Janeiro, RJ, Brazil.  \\
  \inst{2} Departamento de Matem\'atica Aplicada,
Universidade do Estado do Rio de Janeiro, Rua S\~ao Francisco Xavier 524,
20550-013,  Rio de Janeiro, RJ, Brazil.
}
\pacs{05.70.Ln}{Nonequilibrium and irreversible thermodynamics}
\pacs{05.40.-a}{Fluctuation phenomena, random processes, noise, and Brownian motion}
\pacs{05.20.-y}{Classical statistical mechanics}

\abstract{
We present a path integral formalism to compute  potentials for nonequilibrium steady states, reached by a multiplicative stochastic dynamics. We develop a  weak-noise expansion, which allows the explicit evaluation of the potential  in arbitrary dimensions and for any stochastic prescription. We apply this general formalism to study noise-induced phase transitions. We focus on a class of  multiplicative  stochastic lattice models and  compute the steady state phase diagram in terms of the noise intensity and the lattice coupling. We obtain, under appropriate conditions, an ordered phase induced by noise.  By computing entropy production, we show that  microscopic irreversibility is a necessary condition to develop noise-induced phase transitions. This property  of the nonequilibrium stationary state has no relation with the initial stages of the dynamical evolution,  in contrast with previous interpretations, based on  the short-time evolution of the order parameter.
}

\maketitle

\section{Introduction}
Nonequilibrium statistical mechanics is at the stem of  important physical phenomena, many of them at the border with other sciences such as biology, chemistry, geology and even social sciences.
Differently from equilibrium statistical mechanics, there is no closed theoretical framework to deal with out-of-equilibrium systems, being the theory of stochastic processes~\cite{vanKampen} one of the natural approaches to describe them. 
These systems have been traditionally  modeled by stochastic differential equations, as well as through  Fokker-Planck equations. Moreover, functional path integral approaches have also been introduced~\cite{WioBook2013}. 
The latter approach is more adaptive to explore symmetries and general formal aspects of stochastic dynamics, such as out-of-equilibrium fluctuations theorems~\cite{Kurchan1998,Corberi2007, CamillePreprint}.  

Recently, clear advances in the path integral approach of  multiplicative processes were made~\cite{arenas2010,arenas2012,Arenas2012-2,Miguel2015}.  Multiplicative noise naturally describes inhomogeneous diffusion in which fluctuations depend on the state of the system. Concrete applications are very diverse,  covering a  broad range of interest as, for instance,  micromagnetism~\cite{Aron2014}  and  early life biology~\cite{Goldenfeld2015}. 
On the other hand, extended multiplicative systems present  challenging phenomena, such as noise-induced phase transitions (NIPT), stochastic resonance and pattern formation~\cite{SanchoBook,DickmanBook2005}. 
Although it is possible to define order parameters and  susceptibilities in nonequilibrium steady states, the classification of phase transitions in universality classes is still underdeveloped. In fact, a complete general theory of  the nonequilibrium Renormalization Group is still lacking. The definition and evaluation of thermodynamical potentials, such as the Helmholtz or Gibbs free energies, are not completely developed in out-of-equilibrium statistical mechanics.
 Indeed, there is an actual discussion about the possibility of having a thermodynamic description of nonequilibrium steady states~\cite{DickmanPreprint2015},  although some steps forward for par\-ti\-cular systems have been achieved~\cite{Graham1998,WioBook2012,Wio2002,Wio2007,vanWijland2007,Parrondo2015}.

The main goal of this letter is to provide a general formalism to compute potentials for describing  nonequilibrium stationary states reached by a multiplicative Langevin dynamics. Based on a recently introduced path integral formalism~\cite{Miguel2015}, we present  nonequilibrium generating functionals, appropriated to compute  order parameter correlations.  We also introduce a systematic and controlled non-perturbative weak-noise approximation to compute these potentials. 

We explicitly compute, in the Gaussian fluctuation approximation, the nonequilibrium potential for a wide class of Langevin dynamics, in arbitrary spatial dimensions and for any stochastic prescription, including  It\^o, Stratonovich and anti-It\^o prescriptions. 
This last point is essential in the study of entropy production and fluctuation theorems, since time-reversal transformations generally mix different prescriptions~\cite{Arenas2012-2}.  As a concrete example, we apply the formalism to a set of  models,  including  the simplest lattice model proposed by Van den Broeck, Parrondo and Toral (VPT)~\cite{Parrondo1994} and   a related continuum model of Genovese, Mu\~noz and Sancho~\cite{Sancho1998}.
We show that our approximation procedure correctly captures the physics of NIPT and, differently from equilibrium phase transitions, we establish a deep connection between NIPT and microscopic irreversibility~\cite{crooks2000}.   

\section{ Dynamical generating functionals}
 We begin by considering a system of Langevin equations given by
\begin{equation}
 \frac{dx_i(t)}{dt} = f_i({\bf x}(t)) + 
g_{ij}({\bf x}(t))\eta_j(t)   \; , 
\label{eq:LangevSystem}
\end{equation}
 where  $i = 1,\ldots,n$, $j = 1,\ldots,m$, ${\bf x} \in \Re^n$ and  $\eta_j(t)$ are $m$ independent Gaussian white noises:
$\left\langle \eta_i(t)\right\rangle   = 0$, 
$\left\langle  \eta_i(t), \eta_j(t')\right\rangle =\sigma^2 \delta_{ij} \delta(t-t')$, where $\sigma^2$ measures the noise intensity.
We use bold face characters for vector variables and  summation over repeated indices is understood. 
The drift force $f_i({\bf x})$ and  the diffusion matrix $g_{ij}({\bf x})$ are,  in principle,  arbitrary smooth functions of ${\bf x}(t)$. Along this letter, we use the Generalized Stratonovich prescription, parametrized by a real number  $0\le \alpha\le 1$~\cite{Arenas2012-2}.  Time correlation functions can be computed by performing functional derivatives of a  generating functional written in terms of functional integrals. Functional integral techniques in different discretization prescriptions have been developed in Refs.~\cite{Langouche1979,TirapeguiBook1982,Lubensky2007}. We have used a generalization of the Martin-Siggia-Rose-Janssen-deDominicis formalism~\cite{MSR1973,Janssen1976,deDominicis}, that we have recently implemented~\cite{Miguel2015} to deal with the  multi-variable dynamics of Eq.~(\ref{eq:LangevSystem}) in the $\alpha$-prescription. The generating functional can be wri\-tten, after integrating out auxiliary variables,  in terms of a functional integral over the vector variable ${\bf x}(t)$,   
\begin{equation}
Z[{\bf J}]=\int{\cal D}{\bf x}\; {\det}^{-1}(g)\; e^{-\frac{1}{\sigma^2}\left\{S[{\bf x}]-\int_{-\infty}^{\infty}dt' {\bf J}\cdot{\bf x}(t')\right\}} \; ,
\label{eq:ZS}
\end{equation}
where the ``action'' is given by
\begin{eqnarray}
\lefteqn{
S  =  \int_{t_i}^{t_f}   \!\! dt \;\left\{  \frac{1}{2}  
 \left[ \dot x_\ell - \Gamma_\ell \right] 
 [g^2]^{-1}_{\ell m}
 \left[ \dot x_m - \Gamma_m \right] + \alpha\sigma^2 \partial_k f_k \right.} \label{eq:action-alpha} \\
&+& \left.\frac{1}{2} \alpha^2 \sigma^2\left[ \partial_m g_{kj}(x) \partial_k g_{mj}(x) - \partial_m g_{mj}(x) \partial_i g_{ij}(x) \right]  \right\}   ,
 \nonumber 
\end{eqnarray}
with $\Gamma_\ell({\bf x})=f_\ell({\bf x}) - \alpha\sigma^2 g_{\ell j}({\bf x})\partial_i g_{ij}({\bf x})$.
The second and third terms of Eq.~(\ref{eq:action-alpha}), proportional to $\alpha$ and $\alpha^2$ respectively, comes from the non-trivial Jacobian associated with  the change of variables 
 $\eta_i(t)\to x_i(t)$. These terms, together with  $\alpha-$calculus rules~\cite{footnote} are essential to implement any consistent approximation procedure.
In Eq.~(\ref{eq:ZS}),  ${\bf J}(t)$ is a vectorial source necessary to compute correlation functions.  

Formally,  $Z[{\bf J}(t)]$  plays the same role as the partition function in equilibrium statistical mechanics. Then, it is immediate to define the functional  $F[{\bf J}(t)]=-\sigma^2\ln Z[{\bf J}]$, from which we can
compute  a ``local order parameter'' 
\begin{equation}
M_i(t)\equiv \langle x_i(t)\rangle=-\frac{\delta F[{\bf J}]}{\delta J_i(t)}\; .
\label{eq:deltaF}
\end{equation}
The dynamical variable  ${\bf M}(t)$, in principle, is not an order parameter. However, as we will show below, the homogeneous long-time limit behaves like an actual order parameter, detecting order-disorder phase transitions. 

In order to define a generating  functional in terms of the order parameter, we perform a Legendre transformation in the following way, 
\begin{equation}
G[{\bf M}(t)]=\int dt\; {\bf J}(t)\cdot {\bf M}(t)+F[{\bf J}(t)]\; ,
\label{eq:Legendre}
\end{equation} 
where ${\bf J}(t)\equiv {\bf J}[{\bf M}(t)]$ is defined by inverting  Eq.~(\ref{eq:deltaF}).
It is immediate to verify
$\delta G[{\bf M}]/\delta M_i(t)=J_i(t)$, that can be interpreted as  the nonequilibrium state equation.

Assuming that, at long times, the system reaches an homogeneous stationary state, we can define the nonequilibrium potential as
\begin{equation}
G_{\rm st}(M)=\lim_{t\to \infty}\lim_{N\to \infty}\frac{1}{N} G[{\bf M}(t)]\; ,
\label{eq:Gst}
\end{equation}
where $N$ is the number of degrees of freedom and the order parameter  
$M=(1/N)\lim_{t\to \infty} \sum_{i=1}^{N} \langle x_i(t)\rangle$.

The expressions in Eqs.~(\ref{eq:Legendre}) and~(\ref{eq:Gst}) are the main proposals of this letter. The potential $G_{\rm st}(M)$  is the analog of the  Gibbs free energy for  describing nonequilibrium stationary states. The critical behavior of the nonequilibrium state is codified in the analytical properties of $G_{\rm st}(M)$. 

\section{ Weak noise expansion}
Due to the factor $1/\sigma^2$ in the exponential of  Eq.~(\ref{eq:ZS}),  the functional integral can be computed in the saddle-point  plus Gaussian fluctuations approximation, for small $\sigma$. Assuming there is only one trajectory $x^0_i(t)$ that extremize  the action,  we decompose the integration variables as,  
 \begin{equation}
x_i(t)=x^{0}_{i}(t) + \delta x_{i}(t)\; ,
\label{eq:x0+deltax}
\end{equation}
where $x^{0}_{i}(t)$ is a solution of 
\begin{equation}
\left. \frac{\delta S[{\bf x}]}{\delta x_i(t)}\right|_{{\bf x}(t)={\bf x}^0}=J_i(t)
\label{eq:x0}
\end{equation}
and  $\delta x_{i}(t)$ represent small fluctuations.
Replacing Eq.~(\ref{eq:x0+deltax}) into Eq.~(\ref{eq:ZS}) and expanding in powers of $\delta x_{i}(t)$ up to quadratic order, we find, after the Gaussian integration, 
\begin{equation}
F[{\bf J}]= S[{\bf x}^{0}]- \int dt \;{\bf J}\cdot {\bf x}^{o}
+\frac{\sigma^2}{2} {\rm Tr}{\ln} [{\bf S}^{(2)}]+\ldots \; ,
\label{eq:Ffl}
\end{equation}
where the components of the fluctuation matrix ${\bf S}^{(2)}$ are given by
\begin{equation}
S^{(2)}_{ij}(t,t') = \frac{\delta^{2}S[{\bf x}]}{\delta x_{i}(t)\delta x_{j}(t')}\Bigg \vert_{{\bf {\bf x}(t)}={\bf x}^{0}(t)}, \label{eq:propagator} 
\end{equation} 
and the ellipsis represents order $\sigma^4$ terms. 
Using Eqs.~(\ref{eq:deltaF}) and~(\ref{eq:Ffl}), we find for the order parameter,
\begin{equation}
M_i(t)=
x^{0}_{i}(t) - \frac{\sigma^2}{2}{\rm Tr}\left\{ [{\bf S}^{(2)}]^{-1} \frac{\delta{\bf S}^{(2)}}{\delta J_i(t)}\right\}.
\label{eq:Mfl}
\end{equation}
For $\sigma\to 0$, the order parameter is essentially the ``classical'' solution ${\bf M}(t)={\bf x}^0(t)$, obtained by solving Eq.~(\ref{eq:x0}). Fluctuations change this result in a non-trivial way. To compute  the Legendre transformation of Eq.~(\ref{eq:Legendre}),   we invert  Eq.~(\ref{eq:Mfl}) perturbatively in powers of $\sigma^2$. Retaining the leading order terms, we find the expression
\begin{equation}
G[{\bf M}(t)] =  S[{\bf M}(t)]+ \frac{\sigma^2}{2}{\rm Tr}\ln \left\{{\bf S}^{(2)}[{\bf M}(t)]\right\} \; .
\label{eq:Gfl}
\end{equation}
Eq.~(\ref{eq:Gfl}) is the explicit expression of the generating functional of the local order parameter in the Gaussian fluctuations approximation. This general result allows the  computation of  the time-dependent order parameter by  solving the  dynamical set of equations  $\delta G[{\bf M}]/\delta M_i(t)=0$, with $i=1,\ldots, N$.  
At this point, it is important to stress that Eq.~(\ref{eq:Gfl}) was computed assuming that there is only one path $x_0(t)$ which solves Eq.~(\ref{eq:x0}). In the case of multiple solutions, the method should be generalized by computing Gaussian fluctuations around each solution and summing up each contribution~\cite{WioBook2013}.

\section{ Lattice models}
Let us compute $G[{\bf M}(t)]$  for a model consisting in a set of $N$ stochastic variables whose dynamics are driven by the same drift $f(x)$ and the same diffusion function $g(x)$. Each degree of freedom is  arranged in a $d$-dimensional hypercubic lattice and we consider short-range lattice couplings. Then, in Eq.~(\ref{eq:LangevSystem}), we will take $f_i({\bf x})=f(x_i)+F_i({\bf x})$ and $g_{ij}({\bf x})=g(x_i)\delta_{ij}$, where $F_i({\bf x})$ represent the lattice couplings.
In the absence of couplings, $F_i({\bf x})=0$, the system converges, at long times, to an equilibrium state. Imposing the Einstein relation  $f(x)=-(1/2)g(x)^2 dV(x)/dx$, where $V(x)$ is a classical potential, the equilibrium distribution is $P_{\rm eq}\sim  \exp\{-(1/\sigma^2) U_{\rm eq}(x)\}$,
with the  potential $U_{\rm eq}(x)=V(x)+(1-\alpha)\ln g^2(x)$~\cite{Arenas2012-2}.
The equilibrium potential depends on the stochastic prescription $\alpha$ and,  in general, it is not of the Boltzmann type, except for $\alpha=1$. Considering even potentials, $V(x)=V(-x)$, with a single minimum at $x=0$ and $g''(0)>0$, we have $\langle x\rangle=0$. This behaviour can change completely in the presence of lattice couplings. Let us consider, for instance,  the simplest lattice interaction, 
\begin{equation}
F_i({\bf x})=\left(\frac{D}{2d}\right) \sum_{x_j\in n(x_i)} \left(x_j-x_i\right)\; ,
\end{equation}
where $n(x_i)$ denotes the set of first neighbors of $x_i$,   $D$ is the coupling constant and  $d$ is the number of dimensions of the hypercubic lattice.
Due to interactions, the fluctuation  matrix  $S^{(2)}_{ij}(t,t')$ is not diagonal, neither in time, nor in lattice indexes. For this reason, ${\rm Tr}\ln {\bf S}^{(2)}$ in Eq.~(\ref{eq:Gfl}) is a cumbersome evaluation.
Fortunately, in the stationary and homogeneous limit the coefficients are constants and  it is possible to   diagonalize ${\bf S}^{(2)}$ by Fourier transforming  in time and  lattice indexes. 
While the frequency spectrum is continuos, the wave-vector modes are discrete for finite systems\cite{Appert-Rolland2008}. Since we are interest in looking for phase transitions, we consider an infinite system by making the limit $N\to \infty$. In this case, the wave-vector spectrum is continuous and the Fourier transform takes the general form $S^{(2)}(\omega,\cos({\vec k}\cdot {\vec a}))$, where  ${|\vec a|}$ is the lattice  constant. In this representation, the trace should be computed by integrating $\ln S^{(2)}$ over $\omega$ and ${\vec k}$ in the first Brillouin zone of the corresponding dual lattice. Provided we are interested  in the long distance behavior of the stationary state, we can further simplify this expression by taking the limit (hydrodynamic regime) $|{\vec k} \cdot  {\vec a}|<<1$. This approximation naturally introduces an ultraviolet momentum cut-off proportional to  $1/a$. Universal quantities should not depend on the specific value of the cut-off. However, we expect that non-universal features, such as the critical noise, should, in general, depend  on it. In this limit, the fluctuation kernel takes the simpler form  
$S^{(2)}(\omega,\vec k)=\omega^2+A |\vec k|^2+\Sigma$, where $A$
and $\Sigma$ are constants. Then,  we can finally  write the nonequilibrium  potential as, 
\begin{equation}
G_{\rm st}(M)= S_{\rm st}(M)+\frac{\sigma^2}{2}\int\frac{d\omega}{2\pi}\frac{ d^dk}{(2\pi)^d}
\ln\left(\omega^2+A |\vec k|^2+\Sigma\right).
\label{eq:Gstapprox}
\end{equation}
where the coefficients $A(M)$ and $\Sigma(M)$ are functions of the order parameter and can be computed from the local properties of $V(x)$ and the diffusion function $g(x)$, near $x=0$. 
It is important to emphasize the applicability range  of  
Eq.~(\ref{eq:Gstapprox}). $G_{\rm st}(M)$ is the nonequilibrium potential of the stationary state, reached by the multiplicative Langevin dynamics,  provided this is an homogeneous state. 
To study inhomogeneities, we should define a continuous local order parameter in the thermodynamic limit $M(x)$ and make a gradient expansion to improve  Eq.~(\ref{eq:Gstapprox}).

To explore possible phase transitions, we compute the inverse susceptibility in the disordered phase, 
\begin{equation}
\chi^{-1}_0=\left. \frac{d^2G_{\rm st}(M)}{dM^2}\right|_{M=0}\; .
\end{equation}

To be specific, let us consider the VPT model, where we  choose an harmonic oscillator potential with natural frequency  $\Omega$,  $V(x)=\Omega^2 x^2$, and a diffusion function $g(x)=1+x^2$. 
Computing the coefficients $A$ and $\Sigma$ and exactly integrating out the frequency, the inverse susceptibility takes the form 
\begin{eqnarray}
\chi_0^{-1}&=&\frac{\Omega^2}{2}\left(1+2\tilde\sigma^2\right)
\left\{1+\frac{7}{8\pi}\left(\frac{d}{D}\right) \sigma^2(1+\tilde\sigma^2)\right.\times \nonumber \\
&\times&\left.\int_0^{\Lambda(\sigma)} dx\;
\frac{x^{d-1} }{\sqrt{1+x^2}}\left(1+\frac{1}{28}\frac{1-\tilde\sigma^2}{1+\tilde\sigma^2} x^2\right) 
\right\},
\label{eq:chi}
\end{eqnarray}
where $\tilde\sigma^2=2 (1-\alpha)\sigma^2/\Omega^2$ and  the cut-off  $\Lambda(\sigma)=(\pi/a)(2D/d(1+2\tilde\sigma^2))^{1/2}$. 
The  integral  in Eq.~(\ref{eq:chi}) can be done in terms of hypergeometric functions. We observe that the existence of NIPT is determined by the integral in the second line of Eq.~(\ref{eq:chi}), since the other terms are positive definite. Interestingly, this term comes from fluctuations. Thus, the physics of NIPT is correctly captured by Gaussian fluctuations. It is interesting to note, that the very existence of the NIPT does not depend on the specific value of the ultraviolet cut-off. On the other hand, the position of the critical line is, in general,  cut-off dependent.

\begin{figure}[htb!]
\subfigure[\  $d=2$, $\Omega=1$. Continuous, dashed and dot-dashed lines correspond to $\alpha=0,1/2, 0.8$, respectively.]
{\includegraphics[scale=0.41]{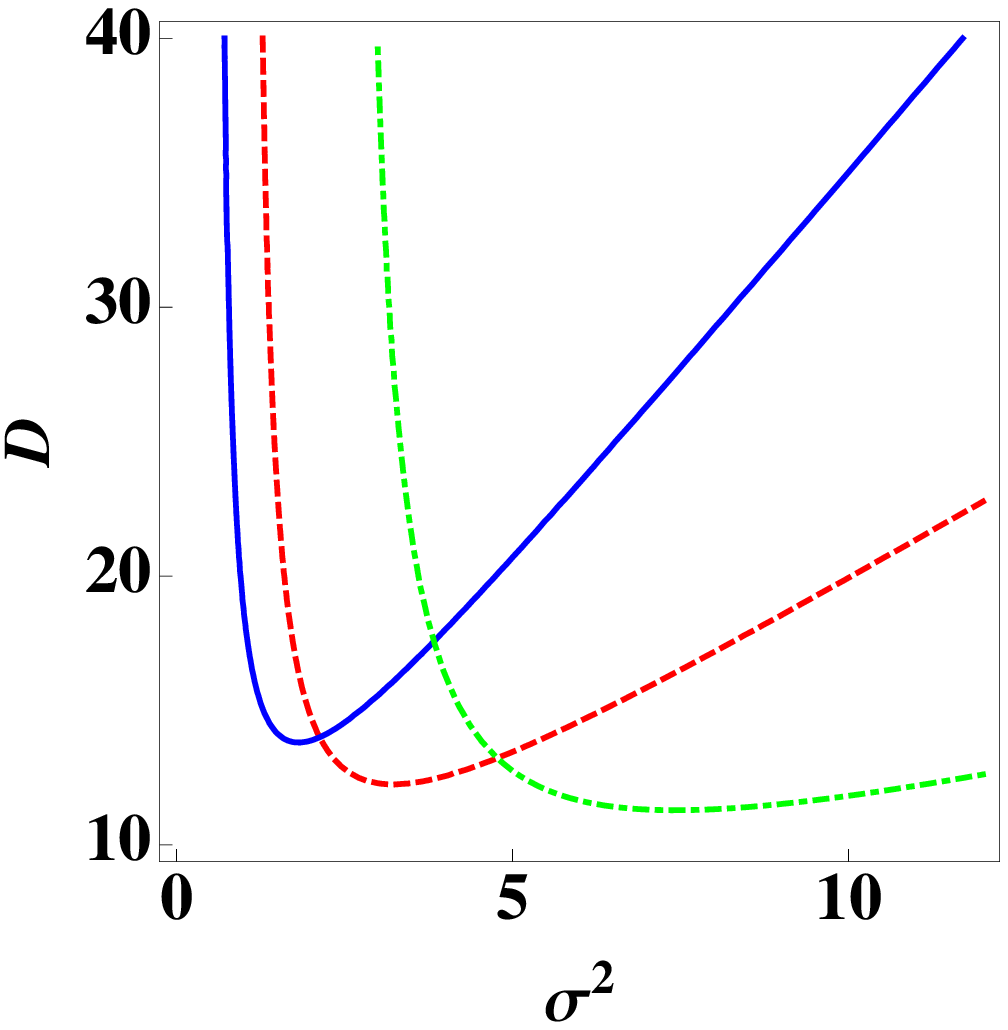}}
\subfigure[\  $\alpha=0$, $\Omega=1$. Continuous, dashed and dot-dashed lines correspond to $d=2,4,6$, respectively.]
{\includegraphics[scale=0.41]{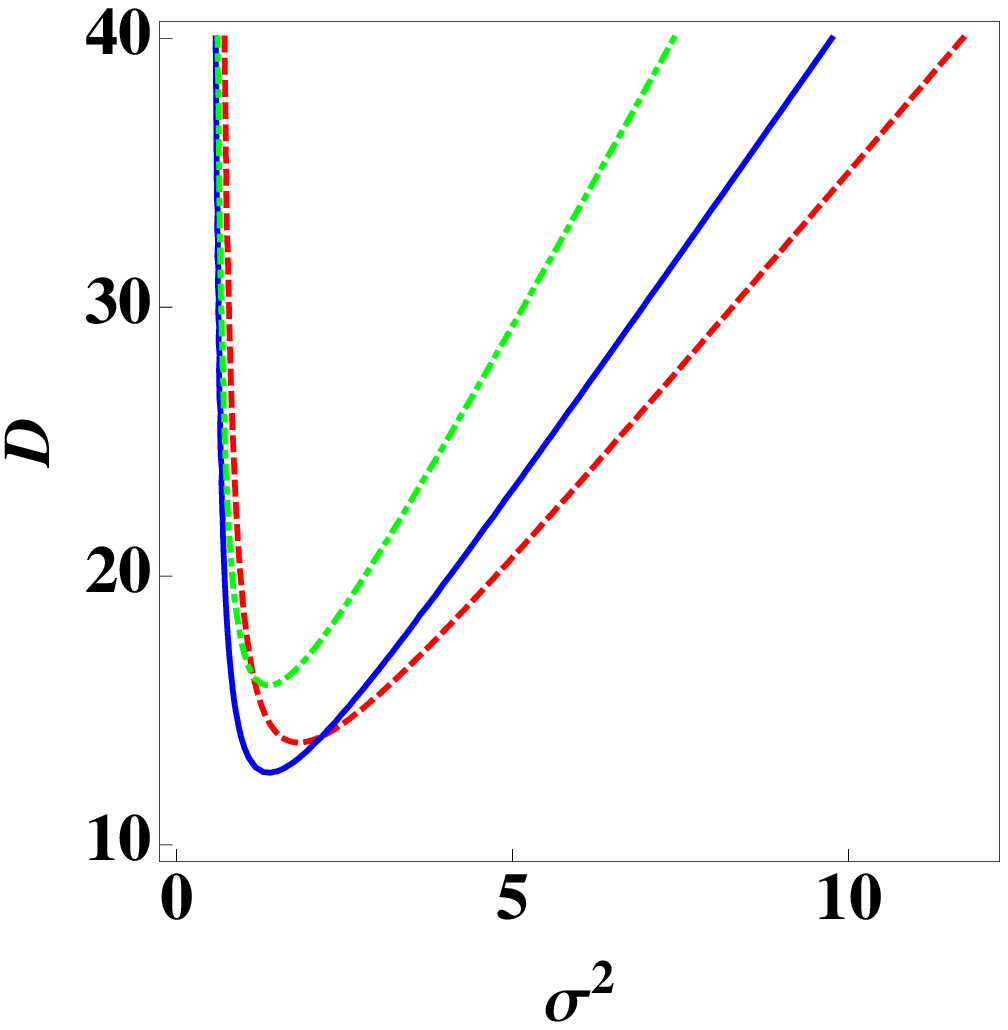}}
\caption{Phase diagram. Critical lines are defined by $\chi_0^{-1}(\sigma^2, D)=0$, with $\chi_0^{-1}$ given by Eq.~(\ref{eq:chi}).  In the exterior part of each line the system is disordered, $M=0$, while in the interior, $M\neq 0$.}
\label{fig:PD}
\end{figure}

The critical line is defined by $\chi^{-1}_0(\sigma^2, D)=0$. 
We depict the results for different values of the parameters in Figure~(\ref{fig:PD}). 
In Figure~(\ref{fig:PD}a) we plot the critical line in $d=2$, for different values of the stochastic prescription. Above a minimum lattice coupling $D_{\rm min}$, we found two continuum phase transitions. For very weak noise the system is disordered, $M=0$. At a  threshold  $\sigma^2_{c1}$, the system orders, $M\neq 0$, breaking in this way  $Z_2$ symmetry. For higher values of the noise, $\sigma^2_{c2}>\sigma^2_{c1}$ it gets disordered  again. 
Comparing the curve $\alpha=1/2$ with the results of Ref.~\cite{Parrondo1997}, we conclude that our procedure correctly describes the behavior of $\sigma_{c1}$. In addition,  the extension of the ordered region $(\sigma^2_{c1}-\sigma^2_{c2})$ is much more accurate  than usual mean-field approximations. We observe that the ordered area in the $D-\sigma^2$ plane growths with $\alpha$. On the other hand, $\sigma^2_{c1}$ as well as $\sigma^2_{c2}$  are increasing functions of $\alpha$, making the phase transition harder to reach. In fact, in the anti-It\^o prescription, $\alpha=1$,  $\chi_0^{-1}$ is positive definite and there is no phase transition. In Figure~(\ref{fig:PD}b), we  depict the critical line in the It\^o ($\alpha=0$) prescription for different values of the spatial dimension $d$. We observe that  $\sigma^2_{c1}$ is very weakly dependent on  $d$. In fact, for $D\to \infty$, the critical noise rapidly converges to  $\sigma^2_{c1}=\Omega^2/[2(1-\alpha)]$ for any value of $d$.  Conversely, $\sigma^2_{c2}$ is strongly dependent  on dimensionality. Form the integral in Eq.~(\ref{eq:chi}), we can see that  the dimensionality essentially enter the cut-off as $\Lambda\sim \sqrt{1/d}$. Thus, we can infer the  following features on the dependence on the position of the critical line with the cut-off: for greater values of $\Lambda$, the minimum coupling constant $D_{\rm min}$ gets deeper, making easer to reach the phase transition. On the contrary for smaller cut-off,  $D_{\rm min}$ rise. In the  strongly coupled regime  $D> D_{\rm min}$, the threshold for NIPT, $\sigma^2_{c1}$, is quite universal, in the sense that is very weakly dependent on the cut-off (in the same way that it is almost independent on $d$, as can be seen from Figure (\ref{fig:PD}b)). However, the  ``normal'' phase transition, $\sigma^2_{c2}$,  has strongly non-universal behavior.

\section{ Time reversal and entropy production}

It is interesting to characterize the nonequilibrium steady state from the point of view of stochastic thermodynamics~\cite{seifert2008}. From this perspective, the concept of time reversal stochastic process is essential. For instance, it can be proved that the time reversed evolution of  a Markov diffusion process described by an It\^o stochastic differential equation is also a Markov diffusion process, however with a different drift~\cite{Haussmann1986,Millet1989}. Moreover,  we have recently showed~\cite{Miguel2015} that a time reversal transformation of a stochastic process defined by a stochastic differential equation in  arbitrary prescription $\alpha$, is  perfectly well defined  and is given by 
\begin{equation}
{\cal T} = \left\{
\begin{array}{lcl}
{\bf x}(t) &\to & {\bf x}(-t)   \\   & & \\
\alpha &\to & (1-\alpha) \\ & & \\
f_i &\to&  f_i  +\left(2\alpha-1\right)\; g_{k\ell}\partial_k g_{i\ell}
\end{array}
\right.
\label{eq:TimeReversal}
\end{equation}
 Note that time reversal mixes different prescriptions and, for this reason, it is important to have a formalism that can deal with all prescriptions consistently.
 With this definition, Crooks relations~\cite{crooks2000} of microscopic reversibility are satisfied in an equilibrium state  reached by a multiplicative Langevin dynamics~\cite{Arenas2012-2}. Conversely, in a nonequilibrium steady state, such as the one we are studying in this letter, microscopic reversibility is broken and there is an entropy production characterizing this state. 

The increase of entropy in the medium, associated with each individual stochastic trajectory~\cite{seifert2008,Toral2015},  can be  defined by  $\Delta s_m=S[{\bf x}]-{\cal T}S[{\bf x}]$,
where ${\cal T} S[{\bf x}]$ is the action of the time reversed process~\cite{Arenas2012-2,Miguel2015}.  For an equilibrium state, the stochastic entropy is a state function,  
$\Delta s_m=\Delta U_{\rm }=U_{\rm eq}({\bf x}_f)-U_{\rm eq}({\bf x}_i)$, where $U_{\rm eq}$ is the equilibrium potential. This is a direct consequence of microscopic reversibility.  
Explicitly computing $\Delta s_m$ for  the VPT model, we find
\begin{equation}
\Delta s_m=\Delta U[{\bf x}] +\sum_k\int_{t_i}^{t_f} dt\; \dot x_k \left(\frac{g'_k}{g_k^3}\right)W_{\rm os}[{\bf x}]
\label{eq:entropy}
\end{equation}
where $W_{\rm os}({\bf x})=(D/d) \sum_{x_j\in n(x_i)} \left(x_j-x_i\right)^2$ is the potential energy of the lattice coupling and $U(x)=V(x)+W_{\rm os}({\bf x})+(1-\alpha)\ln g^2(x)$.  
Clearly, $\Delta s_m$ is not a state function since it depends on the trajectory. The reason behind this behavior is that, in the presence of multiplicative noise, the lattice coupling of the VPT model breaks the Einstein condition, and the stationary state is microscopically irreversible. 
We argue that microscopic irreversibility and, consequently, the breakdown of detailed-balance, is the main cause of noise-induced phase transitions. 
To support that, let us analyze two illustrative examples. It is known that in an additive process, $g'_k=0$,  there is no noise-induced phase transition.  On the other hand, the last term of  Eq.~(\ref{eq:entropy}) is zero and $\Delta s_m$ is a state function. In other words, in the additive noise case, the steady state is an equilibrium one, detailed balance is satisfied, and no NIPT can take place.    
To stronger support this point of view, let us consider a true multiplicative process with a  ``slightly'' modified  lattice coupling, 
\begin{equation}
F_i({\bf x})=\left(\frac{D}{2d}\right)  g^2(x_i)\sum_{x_j\in n(x_i)} \left(x_j-x_i\right)\;.
\end{equation}
This is an harmonic first neighbor interaction locally weighted by the function $g^2(x_i)$.  
This coupling satisfies  Einstein relation since the total drift force can now  be written as $f_k({\bf x})= -(1/2)g_k^2\partial_k U(\bf x)$. Consequently, the associated entropy is $\Delta s_m=U({\bf x}_f)-U({\bf x}_i)$, indicating that this system is microscopically reversible. Interestingly, by  computing the potential $G_{\rm st}(M)$ and  the inverse susceptibility  $\chi^{-1}_0$,  we verify that the latter is positive definite, implying that  there is no phase transition in this model. 

These facts strongly suggests that {\em microscopic irreversibility of the steady state is a necessary condition for noise-induced phase transitions}.

\section{Conclusions} 
We have presented a path integral formalism to compute potentials for nonequilibrium steady states, reached at long times by  multiplicative Langevin dynamics. 
The formalism is completely general and can be applied to study, for any stochastic prescription, a variety of models  presenting interesting features, such as noise-induced phase transitions, stochastic resonance and  pattern formation.  We have also developed a controlled weak noise expansion which correctly captures fluctuation-induced phenomena. 
In particular, we have analyzed the physics of NIPT by computing the stationary state potential for a general class of lattice models. For the particular VPT model, 
we have verified that the approximation developed, not only  captures the qualitative behavior, but also improves previous estimations for the critical line.   
Finally, we have shown that {\em microscopic irreversibility is a necessary condition for  NIPT phenomena}.  In such cases, the steady state is  characterized by the presence of currents or, equivalently, by a non-zero entropy production rate. This property  of the steady state
 has no relation with the initial stages of the dynamical evolution,  in contrast with other interpretations~\cite{Parrondo1997}, based on  the short-time behavior of the order parameter. 
We believe that the results presented in this letter open many interesting possibilities to advance our understanding of out-of-equilibrium critical phenomena.       

\acknowledgements
We acknowledge Daniel A. Stariolo and Horacio Wio for useful discussions. 
The Brazilian agencies CNPq, FAPERJ and CAPES are acknowledged for
partial financial support. D.G.B.  acknowledges ICPT for a Senior Associate award.


\end{document}